\documentclass{aa}  

\usepackage{graphicx}
\usepackage{amssymb,amsmath}
\usepackage{latexsym}
\usepackage{hyperref}								
\usepackage{lmodern}
\usepackage{mathtools}
\usepackage{epsfig}
\usepackage{color}
\usepackage{cancel}
\usepackage{natbib}
\usepackage{needspace}
\usepackage[varg]{txfonts}

\makeatletter
\renewcommand*\aa@pageof{, page \thepage{} of \pageref*{LastPage}}
\makeatother

\begin{document}
\title{Astrometry, orbit determination, and thermal inertia of the Tianwen-2 target asteroid (469219) Kamo`oalewa}

   \author{Marco Fenucci\inst{1,2}
          \and
          Bojan Novakovi\'{c}\inst{3}
          \and
          Pengfei Zhang\inst{4}
          \and
          Albino Carbognani \inst{5}
          \and
          Marco Micheli \inst{1,6}
          \and
          Laura Faggioli\inst{1,6}
          \and
          Francesco Gianotto\inst{1,6}
          \and 
          Francisco Oca\~{n}a\inst{7,8}
          \and
          Dora F\"{o}hring\inst{1,6}
          \and 
          Juan Luis Cano\inst{9}
          \and
          Luca Conversi\inst{1}
          \and 
          Richard Moissl\inst{1}
          }
          
   \authorrunning{M. Fenucci et al.}
    \institute{ESA ESRIN / PDO / NEO Coordination Centre, Largo Galileo Galilei, 1, 00044 Frascati (RM), Italy \\ \email{marco.fenucci@ext.esa.int}
    \and
    Deimos Italia s.r.l., Via Alcide De Gasperi, 24, 28060 San Pietro Mosezzo (NO), Italy
    \and
    Department of Astronomy, Faculty of Mathematics, University of Belgrade, Studentski trg 16, Belgrade, 11000, Serbia
    \and
    Center for Lunar and Planetary Sciences, Institute of Geochemistry, Chinese Academy of Sciences, Guiyang (Guizhou), PR China
    \and
    INAF - Osservatorio di Astrofisica e Scienza dello Spazio, Via Gobetti, 93/3, 40129 Bologna, Italy
    \and
    Starion Italia, Via di Grotte Portella, 28, 00044 Frascati (RM), Italy
    \and
    ESA ESAC / PDO, Bajo del Castillo s/n, 28692 Villafranca del Castillo, Madrid, Spain
    \and
    Deimos Space S.L.U., Ronda de Poniente, 19, 28760 Tres Cantos Madrid, Spain
    \and
    ESA ESOC / PDO, Robert-Bosch-Straße 5, 64293 Darmstadt, Germany
    }

   \date{Received --- / Accepted ---}


  \abstract
     {(469219) Kamo`oalewa is a small near-Earth asteroid, which is currently a quasi-satellite of the Earth. Lightcurve measurements also reveal a rotation period of only about 30 minutes. This asteroid has been selected as the target of the Tianwen-2 sample-return mission of the China National Space Administration.}
     {The first goal of this paper is to observe and improve the orbit determination of (469219) Kamo`oalewa, and better determine the Yarkovsky effect acting on it. The second goal is to estimate the thermal inertia of the asteroid, taking advantage of an improved Yarkovsky effect determination.}
     {Our observational campaign imaged the asteroid from the Loiano Astronomical Station and from the Calar Alto Observatory, in March 2024. We also accurately re-measured a precovery detection from the Sloan Digital Sky Survey from 2004. New astrometry was later used in a 7-dimensional orbit determination, aimed at estimating both the orbital elements and the Yarkovsky effect. Thermal inertia is later studied by using the ASTERIA, a new method that is suitable to estimate thermal inertia of small asteroids.}
     {We detected a semi-major axis drift of $(-67.35 \pm 4.70) \times 10^{-4}$ au My$^{-1}$ due to the Yarkovsky effect, with a high signal-to-noise ratio of 14. The new orbit solution also significantly reduced the position uncertainty for the arrival of the Tianwen-2 spacecraft. 
     By using different models for the physical parameters of Kamo`oalewa, the ASTERIA model estimated the thermal inertia at $\Gamma = 150^{+90}_{-45}$ J m$^{-2}$ K$^{-1}$ s$^{-1/2}$ or $\Gamma = 181^{+95}_{-60}$ J m$^{-2}$ K$^{-1}$ s$^{-1/2}$. 
     }
     {}



   \maketitle

\section{Introduction}
\label{s:intro}

The small near-Earth asteroid (NEA) (469219) Kamo`oalewa\footnote{Previously known also with the provisional designation 2016~HO$_3$.} was discovered on 27 April 2016 by the Panoramic Survey Telescope and Rapid Response System \citep[Pan-STARRS,][]{denneau-etal_2013} in Haleakala, Hawaii. This object attracted the attention of astronomers since its discovery, announced in the Minor Planet Electronic Circular (MPEC) 2016-H63\footnote{\url{https://minorplanetcenter.org/mpec/K16/K16H63.html}}, because of its exceptional dynamical and physical properties. 
Kamo`oalewa is an Earth quasi-satellite, meaning that the relative mean longitude with respect to Earth librates around zero. In more intuitive terms, a quasi-satellite appears to orbit around the Earth in a rotating reference frame in which the position of our planet is fixed, without being gravitationally bound to it. In fact, Kamo`oalewa is placed at about 100 lunar distances, much further than the 3.9 lunar distances corresponding to the Hill sphere of influence of Earth. 
On a timescale of $\sim$100 kyr, the orbit of Kamo`oalewa periodically switches between a quasi-satellite configuration and a horseshoe configuration \citep{delafuente-delafuente_2016}, and it shows chaotic behavior \citep{fenucci-etal_2022}. Numerical simulations over a longer period also showed that it could remain a companion of the Earth for at least the next 500 kyr \citep{fenucci-novakovic_2021}. 
The Yarkovsky effect influences the orbits of asteroids smaller than 30 km in diameter \citep[see e.g.][]{vokrouhlicky-etal_2015}. In recent years, it has been shown that the effect also plays a role in the motion of Kamo`oalewa, though detecting and computing a precise value from the current observations available at the Minor Planet Center (MPC) presents some challenges \citep{liu-etal_2022, hu-etal_2023}.

From the point of view of physical properties, Kamo`oalewa is a triaxial sub-hundred-meter NEA with an estimated size of $69 \times 58 \times 51$ m \citep{zhang-etal_2024}. Multiple lightcurves taken at different times also confirmed a rotation period of only 27.34 minutes \citep{sharkey-etal_2021}, well below the 2.2 h limit under which cohesion-less ruble-pile asteroids undergo disruption \citep{pravec-harris_2000, carbognani_2017}. 
Spectral data in the visible and near-infrared wavelengths were obtained with the Large Binocular Telescope (LBT), and suggested a classification as either an S or L-type, with S-type being the most likely \citep{reddy-etal_2017, zhang-etal_2024}. However, zJHK colors obtained at longer wavelengths do not resemble the behavior of either these two types, while it shows properties similar to those of lunar material. For this reason, and because of the peculiar dynamical properties, an origin as a lunar ejecta was proposed \citep{sharkey-etal_2021, castro-etal_2023}. Further, the Giordano Bruno crater on the Moon was proposed as a possible source of (469219) Kamo`oalewa \cite{jiao-etal_2024}.
In contrast to the mentioned lunar origin explanation, a recent study proposed that the peculiar spectral properties are the result of a long exposition to space-weathering \citep{zhang-etal_2024}. This suggests that Kamo`oalewa originated in the main asteroid belt, and migrated to the near-Earth region as a result of the combined effect of the $\nu_6$ secular resonance with Saturn and the Yarkovsky semi-major axis drift \citep{granvik-etal_2017, granvik-etal_2018}. 
Therefore, a clear answer to the question of the origin of Kamo`oalewa is still a matter of discussion in the community.

Kamo`oalewa is also the target of the Tianwen-2 mission \citep{zhang-etal_2021}, the first sample-return asteroid mission of the China National Space Administration (CNSA). The spacecraft is planned to be launched in 2025 and return a sample of the asteroid by 2027. After returning the sample, the spacecraft will continue its journey for the exploration of comet 311P/PANSTARRS. In-situ explorations are capable of unveiling properties and details of asteroids to an astonishing level, as demonstrated by the NASA OSIRIS-REx \citep{lauretta-etal_2019} and the Jaxa Hayabusa2 \citep{watanabe-etal_2019} missions. Samples returned on Earth are also of fundamental importance, because the material can be analyzed in laboratories to a level of detail that can not be achieved by remote analyses performed by spacecraft. At the same time, prior knowledge of physical properties of the target asteroid are of paramount importance for the planning of sample return missions \citep{murdoch-etal_2021}, because the response to spacecraft operations for the sampling collection may vary depending on the material lying on the surface. One of the most critical parameters is the thermal inertia, which indicates the resistance of a material to temperature changes, and gives information about regolith grain size \citep{gundlach-blum_2013, schorghofer-etal_2024}. It is therefore essential to support sample return missions with remote studies of physical properties of the targets. 

The aim of this paper is twofold. The first goal is to improve the quality of the heliocentric orbit of Kamo`oalewa, providing a reliable and accurate determination of the Yarkovsky effect. This is achieved by performing new astrometric measurements from different locations and by accurate orbit determination. The second goal is to use the new determination of the Yarkovsky effect to derive constraints on the thermal inertia of Kamo`oalewa, and give clues about the material lying on its surface. To this purpose, we apply the Asteroid Thermal Inertia Analyzer \citep[ASTERIA,][]{fenucci-etal_2023b, novakovic-etal_2024} method, a novel approach for NEA thermal inertia estimation. This method has been proven to be a valid alternative to classical thermophysical modeling \citep{novakovic-etal_2024, novakovic-fenucci_2024}, and it is the only available method to constrain thermal inertia of small and fast-rotating asteroids \citep{fenucci-etal_2021, fenucci-etal_2023b} such as Kamo`oalewa.  
   
The paper is organized as follows. In Sec.~\ref{s:observations} we present the results of our observation campaign of Kamo`oalewa. In Sec.~\ref{s:methods} we introduce the methods to determine the orbit, to estimate the thermal inertia, and grain size of regolith. In Sec.~\ref{s:res} we present our results. In Sec.~\ref{s:discussion} we discuss our results. In Sec.~\ref{s:conclusions} we summarize the finding of this paper.

\section{Optical observations of Kamo`oalewa}
\label{s:observations}
\subsection{New observations from 2024}
New observations of Kamo`oalewa, dating to 2024-02-02 and taken from the Kitt Peak Observatory (MPC code 695), were announced in the MPEC 2024-D99\footnote{\url{https://www.minorplanetcenter.net/mpec/K24/K24D99.html}}, when the asteroid was still at a visual magnitude of about 22.8. By including these observations in the orbital fit, the ESA automated procedure for Yarkovsky effect determination \citep{fenucci-etal_2024} already signaled a detection with a signal-to-noise ratio (SNR) of about 3.5. Motivated by this result, we organized an observational follow-up campaign aimed at improving the quality of the orbit.

Additional optical observations of Kamo`oalewa were made from the Loiano Astronomical Station, Italy (MPC code 598). The Loiano Astronomical Station, managed by the Astrophysics and Space Science Observatory of Bologna, is equipped with the \textit{G. D. Cassini} 1.52 m f/4.8 Ritchey-Chr\'etien telescope, the second largest instrument in Italy. The Bologna Faint Object and Spectroscopic  Camera (BFOSC) was attached to the telescope, equipped with a Princeton Instruments EEV $1340 \times 1300$ pixel back-illuminated CCD with 20~$\mu$m pixel size. The plate scale in bin 2 mode was 1.16~arcsec~px$^{-1}$ leading to a field of view of $13.0 \times 12.6$~arcmin. 

\begin{figure*}
    \centering
    \includegraphics[width=0.8\textwidth]{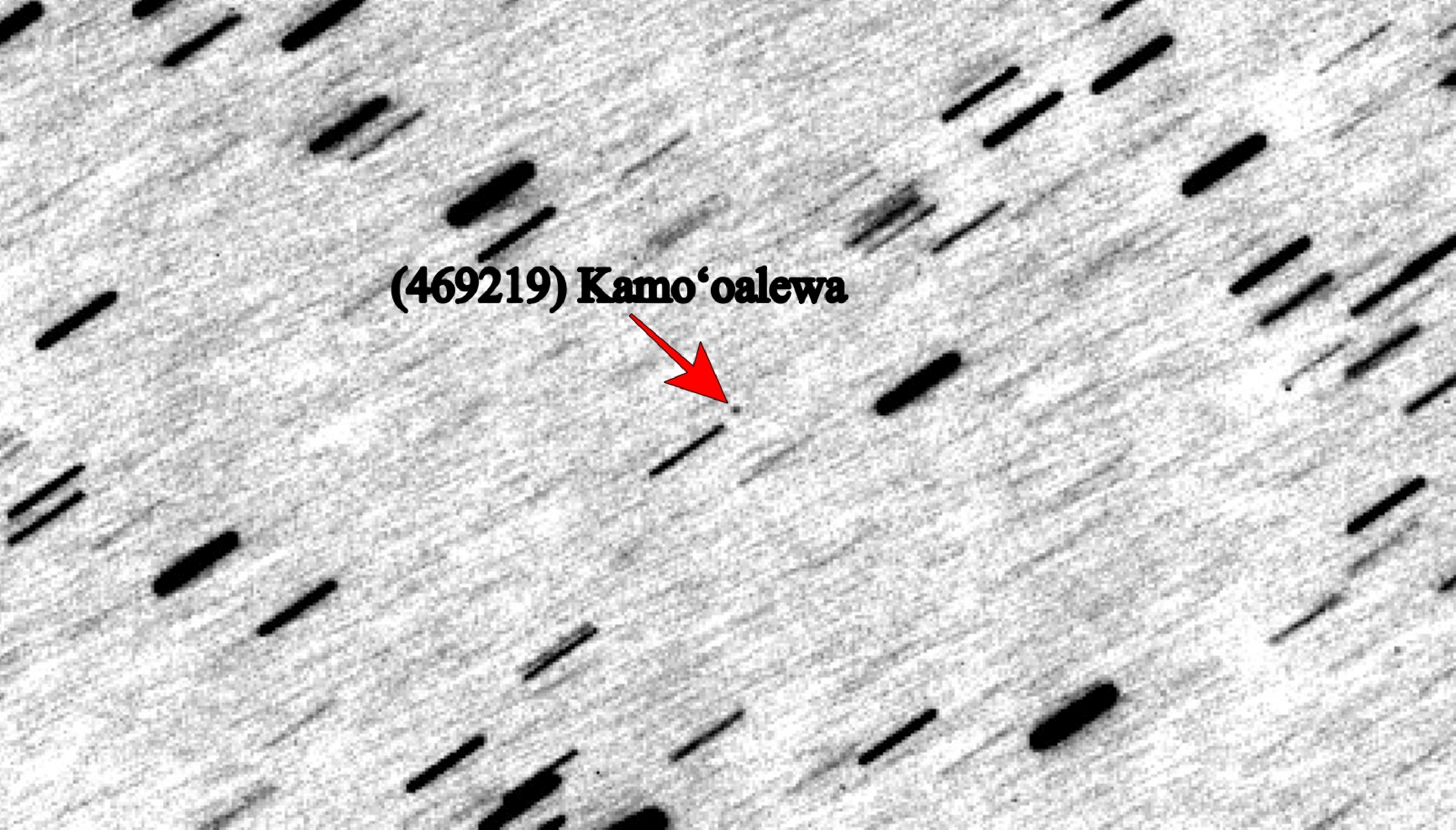}
    \caption{Detection of Kamo`oalewa from the Loiano Astronomical Station (MPC code 598), from 2024-03-07.}
    \label{fig:Kamooalewa_Loaiano}
\end{figure*}

From this station, Kamo`oalewa was observed on 2024-03-05 and 2024-03-07. The sky was clear on the evening of March 5, but the wind was present with gusts of 40-50 km h$^{-1}$. Due to large atmospheric turbulence, the stellar FWHM was 4 arcsec. Kamo`oalewa was imaged from 22:55 to 24:00 UTC in bin 2 mode, without filters and with 60 s exposures. Given that Kamo`oalewa's angular speed was about 1.5 arcsec/minute, this was the maximum time to image the asteroid within a couple of pixels on CCD detector. The asteroid was found using the track and stack technique on sums of 30 images calibrated with master bias and master flat. The SNR of a single short exposure was insufficient to detect it in a single frame, but by shifting successive frames relative to each other and then co-adding the shifted frames, we synthetically created a long-exposure image as if the telescope were tracking the object \citep{Shao_2014}. The plate solveing, track \& stack and astrometric measurements were made using the Astrometrica software \citep{Raab-2012} with the Gaia DR2 star catalog \citep{Lindegren-etal_2018}, and submitted to the MPC. On the evening of 7 March, the sky was still clear but windless, with a better stellar FWHM of 2.5 arcsec, and the asteroid was imaged between 22:50 and 23:32 UTC in the same way as the previous night. The reduced atmospheric turbulence allowed Kamo`oalewa to be found on stacks of only 12 images and with a better SNR than in the previous session. Figure~\ref{fig:Kamooalewa_Loaiano} shows one of the detections made on 7 March. Astrometric positions from the Loiano Astronomical Station were announced in MPEC 2024-E109\footnote{\url{https://www.minorplanetcenter.net/mpec/K24/K24EA9.html}} and MPEC 2024 E-123\footnote{\url{https://www.minorplanetcenter.net/mpec/K24/K24EC3.html}}.

Follow-up of Kamo`oalewa was also obtained with the 0.8 m f/3 Schmidt Telescope (MPC code Z84) installed at the Calar Alto Observatory, Spain, and currently operated by the ESA Planetary Defence Office. The target was observed with a total integration time of 50 minutes, resulting in a single but solid detection of the object. We extracted accurate Gaia-based astrometry of this detection, and the computed formal error bars were smaller than $\pm0.2$ arcseconds in each coordinate.
Astrometric positions obtained in this work are reported in Table~\ref{tab:469219_obs}.

Observations were also attempted on 2024-03-14, 2024-03-15 and 2024-03-16 from the Astronomical Station Vidojevica, Serbia (MPC code C89), which is equipped with the 1.4 m Milankovi\'{c} telescope \citep{vudragovic-etal_2021}. However, this campaign was inconclusive due to bad weather affecting the station location on the selected days.  

After our campaign, observations from the Pan-STARRS2 telescope (MPC code F52) from 2024-03-18 were also reported later in MPEC 2024-F28\footnote{\url{https://www.minorplanetcenter.net/mpec/K24/K24F28.html}}. 

Note that recent optical observations of Kamo`oalewa were performed also by \citep{huang-etal_2024}. However, this new astrometry is not publicly available at the Minor Planet Center (MPC) and therefore not used in this study.

\subsection{Re-measurement of SDSS images}

During the review of the astrometric dataset of Kamo`oalewa, we noticed the possibly crucial role of the two earliest observations available, a pair of precovery detections dating back to 17 March 2004, found in the Sloan Digital Sky Survey \citep[SDSS, MPC code 645,][]{ivezic-etal_2019} image archive. Previous works, performed without the new observations from 2024 described here, showed that the Yarkovsky effect determination is sensitive to these two measurements \citep{hu-etal_2023}. In addition, the astrometry reported to the MPC was of dubious quality. In fact, the star catalog used for the measurement is missing, thus it is difficult to accurately debias the reported astrometry. We therefore proceeded with an accurate re-analysis of these detections. 

SDSS is known for its peculiar observing mode, known as \textit{drift scan}, where each row of the camera is read out at different subsequent times, compensating for the motion of the sky. 
This mode ensures excellent exposure duration accuracy for the photometric purposes of the original survey, but it results in an additional complication when SDSS images are used for astrometric detections, since every line of the image corresponds to a markedly different mid-exposure time, varying by as much as 54 seconds across the field.
Furthermore, the times reported in the headers of SDSS images are not expressed in the usual UTC timescale, but in International Atomic Time (TAI), which at the epoch of these detections ran 32.0 s ahead of UTC. In addition, they also correspond to the readout time of the first row of the camera, which is the end of the exposure for that row. Therefore, the time of the observation must be carefully determined by taking into account all these issues.

The astrometry currently present in the MPC database appears to have been reduced without taking any of these considerations into account. The time tag associated with each astrometric record is simply the TAI time of the end of exposure of the first row of that image. We therefore reviewed these two precovery positions, applying appropriate corrections to UTC, mid-exposure and to the proper time of the row where the object fell. We also re-measured the coordinates of the object itself by using the Gaia DR2 catalog, and computed formal astrometric error bars for each measured coordinate (see Table~\ref{tab:469219_obs}). These positions were used in our analysis, instead of the data originally submitted to the MPC.

We incidentally note that this issue with timing of SDSS astrometric measurements appear to be widespread in the MPC archive, and may need to be separately addressed to ensure appropriate data quality for a large number of objects in their database.

{\renewcommand{\arraystretch}{1.2} 
\begin{table*}[!ht]
    \centering
    \caption{Astrometry of (469219) Kamo`oalewa obtained from the Loiano Astronomical Station (MPC code 598), from the ESA Schmidt Telescope (MPC code Z84), and the re-measurements of images from SDSS (MPC code 645). Columns $\alpha$ and $\delta$ correspond to RA and Dec., respectively. Column $V$ gives the visual magnitude at the time of observation. Columns $\sigma_\alpha$ and $\sigma_\delta$ correspond to the astrometric error in RA and Dec., respectively. }
    \begin{tabular}{lllccll}
        \hline
        \hline
        Time (UTC) & $\alpha$ (deg) & $\delta$ (deg) & $V$ & Station & $\sigma_\alpha$ (arcsec) & $\sigma_\delta$ (arcsec)  \\
        \hline
2024-03-05 23:13:04.6   &211.00194  &51.10909  & 23.6&		598&	0.54  & 0.54\\
2024-03-05 23:42:31.5   &210.98625  &51.11674  & 23.0&    	598&	0.59  & 0.59\\
2024-03-07 22:57:18.6   &209.54914  &51.76714  & 22.3&		598&	0.30  & 0.30\\
2024-03-07 23:14:49.6   &209.53878  &51.7713   & 22.6&		598&	0.30  & 0.30\\
2024-03-07 23:22:51.6   &209.53418  &51.77329  & 21.8&		598&	0.11  & 0.09\\
2024-03-14 00:59:07.22  &204.125784 &53.305496 & 22.6&		Z84&	0.147 & 0.133\\
2004-03-17 11:15:49.76  &234.655345 &27.813972 & 22.9&		645&	0.221 & 0.222\\
2004-03-17 11:20:36.41  &234.654851 &27.814985 & 20.9&		645&	0.193 & 0.193\\
        \hline
    \end{tabular}
    \label{tab:469219_obs}
\end{table*}
}

\section{Methods}
\label{s:methods}
\subsection{Orbit determination}
\label{ss:OD}
Orbit determination is performed with the ESA Aegis orbit determination and impact monitoring software \citep{fenucci-etal_2024b}.
Initial orbital elements are estimated by a least-square procedure aimed at
minimizing the astrometric residuals, defined as the difference between the
observed astrometric positions and the predictions computed by means of a
dynamical model \citep[see, e.g.][]{milani-gronchi_2009}. The force model used
to compute predictions includes the gravitational perturbations of the Sun and
the eight planets from Mercury to Neptune, the Moon, the 16 most massive
main-belt asteroids, and Pluto. The JPL Planetary Ephemeris DE441
\citep{park-etal_2021} is used to retrieve accurate masses of all the massive
perturbing bodies and the positions of the planets and the Moon. The
positions of the 16 main-belt asteroids and Pluto are obtained from precomputed
ephemerides based on our own orbit solutions for these bodies. Complete values
of the masses of the perturbing asteroids can be found in
\citet{fenucci-etal_2024}. Relativistic effects of Sun, Earth, and Moon are
added as a first-order Newtonian expansion \citep{will_1993}, while the
oblateness terms of the Sun and the Earth are taken into account through the
$J_2$ perturbation.
The Yarkovsky effect is expressed as an acceleration along the transverse direction $\hat{\mathbf{t}}$ of the form
\begin{equation}
    \mathbf{a}_t = A_2 \bigg(\frac{1 \text{ au}}{r}\bigg)^2\hat{\mathbf{t}},
    \label{eq:yarko_od_model}
\end{equation}
where $r$ is the distance from the Sun \citep{farnocchia-etal_2013}. The parameter $A_2$ is determined together with the orbital elements in the least-square procedure.

Observational outliers are discarded by using an automated rejection algorithm that computes a value of $\chi^2$ for each observation, and rejects those with a value larger than a threshold $\chi^2_{\text{rej}}$ \citep{carpino-etal_2003}. Star catalog biases are treated with the model by \citet{eggl-etal_2020}, while astrometric errors, when not provided by the observer, are determined through the statistical astrometric error model by \citet{veres-etal_2017}.
When the astrometric errors $\sigma_{\alpha}, \sigma_{\delta}$ for right ascension (RA) and declination (Dec.), respectively, are provided by the observer, a safety term $\sigma_{\alpha, \min}, \sigma_{\delta, \min}$ may be applied to the reported error, correcting them as
\begin{equation}
    \sigma^{\textrm{new}}_{\alpha}= \sqrt{\sigma_{\alpha}^2 + \sigma_{\alpha, \min}^2 },   \quad
    \sigma^{\textrm{new}}_{\delta}= \sqrt{\sigma_{\delta}^2 + \sigma_{\delta, \min}^2 }. 
\end{equation}
Different astrometric measurements are assumed to be uncorrelated, but this may not be true in general \citep{farnocchia-etal_2015b}, especially for observations coming from the same observatory in a short interval of time. To mitigate these correlation effects, we apply a scaling factor of $\sqrt{N/4}$ to the astrometric errors when $N>4$, where $N$ is the number of observations from the same observatory within a tracklet of 8 hours.

\subsection{Thermal inertia estimation}
\subsubsection{The ASTERIA method}
\label{ss:TIest}
Thermal inertia estimation is performed with the ASTERIA method \citep{novakovic-etal_2024}, using the publicly available software \citep{fenucci-etal_2023b}. 
Note that a preliminary thermal inertia estimation of $\Gamma = 402^{+376}_{-194}$ J m$^{-2}$ K$^{-1}$ s$^{-1/2}$ of Kamo`oalewa was obtained by \citet{liu-etal_2024}, with similar methods to those presented here. However, that result was obtained from a Yarkovsky effect determination with low SNR of about 3, and it was affected by considerably large uncertainties. 
The estimation method relies on solving the measured vs. modeled Yarkovsky semi-major axis drift equation
\begin{equation}
\bigg( \frac{\text{d}a}{\text{d}t} \bigg) (a,e,D,\rho,K,C,\gamma,P,\alpha,\varepsilon) = \bigg( \frac{\text{d}a}{\text{d}t} \bigg)_{\text{m}}
    \label{eq:asteria_eq}
\end{equation}
for the thermal conductivity $K$. Other parameters involved in Eq.~\eqref{eq:asteria_eq} are the semi-major axis $a$, the eccentricity $e$, the diameter $D$, the density $\rho$, the heat capacity $C$, the obliquity $\gamma$, the rotation period $P$, the absorption coefficient $\alpha$, and the emissivity $\varepsilon$. 
The left-hand side corresponds to the Yarkovsky semi-major axis drift predicted by a physical model of the asteroid, while the right-hand side $( \text{d}a/\text{d}t )_{\text{m}}$ is the measured drift. It is obtained from the $A_2$ acceleration estimated from orbit determination as
\begin{equation}
    \bigg(\frac{\text{d}a}{\text{d}t}\bigg)_{\text{m}} = \frac{2A_2(1-e^2)}{np^2},
\end{equation}
where $n$ is the mean motion and $p$ is the semi-latus rectum.
Once Eq.~\eqref{eq:asteria_eq} is solved for $K$, the thermal inertia $\Gamma$ is computed as
\begin{equation}
    \Gamma = \sqrt{\rho K C}.
    \label{eq:A2todadt}
\end{equation}

Parameters that are known with a sufficiently small uncertainty, such as the orbital elements, can be assumed to be fixed in the model of Eq.~\eqref{eq:asteria_eq}. On the other hand, unknown parameters are modelled with a suitable distribution. Values and distributions of input parameters have to be chosen ad-hoc, according to the known properties of the object to analyze. The output probability density function for $\Gamma$ is obtained through a Monte Carlo method by randomly choosing the input parameters and solving Eq.~\eqref{eq:asteria_eq} for a large number of combinations of input parameters. 

On the left-hand side of Eq.~\eqref{eq:asteria_eq}, we used a semi-analytical Yarkovsky effect model \citep{vokrouhlicky-etal_2017} that takes into account the effects of orbital eccentricity. The model, however, assumes a spherical asteroid shape and must be adjusted to account for non-sphericity. We incorporated this effect by applying a scaling factor $\xi$ \citep{novakovic-etal_2024} to the measured semi-major axis drift. This factor depends on the asteroid shape \citep{vokrouhlicky_1998}, which is typically inferred from its lightcurve. The correction factor is defined as $\xi = f^{-0.3}$, where $f$ is obtained from the lightcurve amplitude $\Delta m$ as $\Delta m = 2.5 \log_{10}f$.

\subsubsection{Orbital and physical parameters modeling}
To get an estimate as reliable as possible with the ASTERIA method, we used ad-hoc values and distributions of the input parameters needed by the model, according to the current knowledge about orbital and physical properties of Kamo`oalewa. 

The semi-major axis and eccentricity are kept fixed, while the Yarkovsky semi-major axis drift is assumed to be Gaussian distributed, with standard deviation defined by the 1$\sigma$ uncertainty. In addition, the drift was corrected to take into account non-sphericity effects by dividing it by a factor of $\xi = 0.75$ (see Sec.~\ref{ss:TIest}), obtained from the lightcurve amplitude $\Delta m = 1.07$ obtained by \citet{sharkey-etal_2021}. 
{\renewcommand{\arraystretch}{1.2}
\begin{table*}[!ht]
    \centering
    \caption{Different models of physical parameters of Kamo`oalewa used for the estimation of the thermal inertia. }
    \begin{tabular}{ccccc}
         \hline
         \hline
         Parameter & Model 1 & Model 2 & Model 3 & Model 4 \\
         \hline
         Absolute magnitude, $H$         & $24.28 \pm 0.18$         &  $24.28 \pm 0.18$         &  $24.28 \pm 0.18$         &  $24.28 \pm 0.18$               \\
         Albedo, $p_V$                   & $0.1 \pm 0.03$             &  $0.1 \pm 0.03$             &  $0.24 \pm 0.05$            &  $0.24 \pm 0.05$                  \\
         Bulk density, $\rho$            & $2720 \pm 540$ kg m$^{-3}$ &  $2720 \pm 540$ kg m$^{-3}$ &  $2720 \pm 540$ kg m$^{-3}$ &  $2720 \pm 540$ kg m$^{-3}$       \\
         Rotation Period, $P$            & $0.4716 \pm 0.03$ h        &  $0.4716 \pm 0.03$ h        &  $0.4716 \pm 0.03$ h        &  $0.4716 \pm 0.03$ h              \\ 
         Obliquity, $\gamma$             & $107 \pm 10$ deg           &  NEA Population             &  $107 \pm 10$ deg           &  NEA Population                   \\ 
         Heat capacity, $C$              & 800 J kg$^{-1}$ K$^{-1}$   &  800 J kg$^{-1}$ K$^{-1}$   &  800 J kg$^{-1}$ K$^{-1}$   &  800 J kg$^{-1}$ K$^{-1}$         \\ 
         Absorption coefficient, $\alpha$& 0.95                        &  0.95                        &  0.90                        &  0.90                              \\ 
         Emissivity, $\varepsilon$       & 0.984                      &  0.984                      &  0.984                      &  0.984                            \\ 
         Non-sphericity correction factor, $\xi$                           & 0.75 &  0.75                        &  0.75                        &  0.75                              \\ 
         \hline
    \end{tabular}
             \tablefoot{The absolute magnitude is computed by fitting photometric data with an a-priori distribution of $G$ extrapolated from S-type asteroids data. The low albedo value is chosen according to the measured value on LL chondrite powder analogue \citep{zhang-etal_2024}, while the high albedo value is representative of S-type asteroids \citep{marsset-etal_2022}. The bulk density refers to the average of S-type asteroids \citep{carry_2012}. The rotation period is taken from \citet{sharkey-etal_2021}. The value of obliquity is taken from \citet{zhang-etal_2025}, while the NEA population obliquity distribution is taken from \citet{tardioli-etal_2017}. Heat capacity is assumed as a reasonable value for NEAs. The absorption coefficient is obtained by assuming a slope parameter of $G=0.25$ and the assumed albedo. The emissivity corresponds to the average value measured on a large sample of meteorites \citep{ostrowsky-bryson_2019}. The non-sphericity correction factor $\xi$ is obtained from lightcurve amplitude of 1.07 mag \citep{sharkey-etal_2021}.}
    \label{tab:physPar}
\end{table*}
}

We summarize here what is known about the physical parameters of Kamo`oalewa and what we assume about them. The absolute magnitude $H$ is estimated with the Aegis software \citep{fenucci-etal_2024b}, which fits the visual magnitudes reported along with the astrometry by using the $H-G$ model by \citet{bowell-etal_1989}. Visual magnitudes are also debiased with the model by \citet{hoffmann-etal_2024}. 
The absolute magnitude $H$ is fitted by keeping fixed the value of $G$, since there is not enough data to fit both parameters. The slope parameter $G$ depends on physical characteristics \citep[see e.g.][]{veres-etal_2015,shevchenko-etal_2016}, and a-priori distributions can be assumed provided some information about taxonomic type is known. Since the spectrum of (469219) Kamo`oalewa is compatible with that of S-type composition \citep{reddy-etal_2017}, we extracted the distribution of $G$ for S-type asteroids from the Zwicky Transient Facility data \citep{carry-etal_2024}. We then fitted $H$ over a sample of $G$, as outlined in \citep{fenucci-etal_2024c}. This way, we obtained an absolute magnitude of $H = 24.28 \pm 0.18$.

An equivalent diameter of $D = 57$ m was determined by \citep{zhang-etal_2024}, where a geometric albedo (hereafter albedo) of 0.1 was assumed to convert the absolute magnitude to size. This was justified by the fact that the absorption center and spectral slope of laser irradiated LL-chondrite powder at 0.55 $\mu$m that matched those of the spectrum of Kamo`oalewa had an albedo close to 0.1. This albedo value is lower than the typical average albedo of S-type asteroids, and it may be due to space-weathering effects or shock darkening \citep{binzel-etal_2019}. Still, this albedo determination relies on the correlation assumption between irradiated LL-chondrite powder and albedo, which may not be completely reliable. To account for these uncertainties, we produced results for thermal inertia in two cases: a low-albedo case with $p_V = 0.1 \pm 0.03$, and a high-albedo case with $p_V = 0.24 \pm 0.05$ compatible with properties of S-type asteroids \citep[see e.g.][]{marsset-etal_2022}. The diameter distribution is then computed by the well-known conversion formula that uses both the albedo and the absolute magnitude \citep{bowell-etal_1989}. 

Constraints on the bulk density can be set according to properties of S-type asteroids, and a normal distribution with parameters $\rho = 2720$ kg m$^{-3}$ and $\sigma_\rho = 540$ kg m$^{-3}$ was assumed \citep{novakovic-etal_2024} in this work.

The rotation period is well constrained because of the lightcurves obtained by using the Large Binocular Telescope (LBT) and the Discovery Channel Telescope (DCT), and it is considered as a solid determination. In our model, we used the value $P = 0.4716 \pm 0.03$ h as determined by \citet{sharkey-etal_2021}.

A solution of the spin-axis of Kamo`oalewa was also determined from lightcurve inversion \citep{zhang-etal_2025}, at ecliptic longitude and latitude of $133.52 \pm 0.01$ deg and $-10.67 \pm 0.05$ deg, respectively. From these values we could compute an estimated obliquity of 107 deg, to which we added an ad-hoc uncertainty of 10 deg. However, this is the only determination of the obliquity obtained so far for Kamo`oalewa, and therefore we also obtained thermal inertia estimates by assuming the quadratic NEA obliquity distribution by \citet{tardioli-etal_2017}.

Previous results obtained on 2011~PT \citep{fenucci-etal_2021} and on 2016~GE$_1$ \citep{fenucci-etal_2023} showed that the value of heat capacity does not affect the thermal inertia estimates of small super-fast rotators. For this reason, the value of $C$ was kept fixed at 800 J kg$^{-1}$ K$^{-1}$, which is a typical value for NEAs. 
The emissivity was set to the average value of 0.984 obtained from meteorite samples \citep{ostrowsky-bryson_2019}.
The absorption coefficient was computed according to the value of the albedo $p_V$ and the slope parameter $G$ of 0.25, and it corresponds to 0.95 for the low-albedo case, and to 0.90 for the high-albedo case. 
It is worth mentioning that previous works showed that the thermal inertia obtained with this method is not particularly sensitive to the chosen values of $C, \alpha$ and $\varepsilon$ \citep{fenucci-etal_2021, fenucci-etal_2023, novakovic-etal_2024}, as long as they are kept in a range of values reasonable for NEAs.

In summary, we assumed four different models for the physical parameters, differing in the value of the albedo and the assumed distribution of the obliquity. The specific values for each model are reported in Table~\ref{tab:physPar}.

\subsection{Grain size estimation}
\label{ss:grainsize}
Regolith grain size is estimated with the model by \citet{gundlach-blum_2013}. The thermal conductivity of a granular medium at temperature $T$, composed of grains with radius $r$, and having a filling factor $\phi$ is modelled as
\begin{equation}
    \lambda(r,T,\phi) = \lambda_{\text{solid}} H(r,T,\phi) + 8\sigma\varepsilon T^3 \Lambda(r,\phi),
    \label{eq:gundblum}
\end{equation}
where $\sigma$ is the Stefan-Boltzmann constant and $\varepsilon$ is the emissivity of the material. The first term of Eq.~\eqref{eq:gundblum} is the conductive term, describing heat conduction through grains, while the second one is the radiative term, describing the heat radiated across porous spaces.
In addition, in Eq.~\eqref{eq:gundblum} we have
\begin{equation}
    \Lambda(r, \phi) = e_1 \frac{1-\phi}{\phi}r, \quad e_1 = 1.34,
\end{equation}
and $H$ is the dimensionless Hertz factor given by
\begin{equation}
    H(r,T,\phi) = \bigg( \frac{9\pi}{4} \frac{1-\mu^2}{E} \frac{\gamma(T)}{r} \bigg)^{1/3} f_1 e^{f_2\phi}\chi.
\end{equation}
In the above equation, $\mu$ is the Poisson ratio, $E$ is the Young modulus, and $\gamma(T)$ is the specific surface energy of the grain material. The parameters $f_1, f_2$ and $\chi$ are empirically determined by \citet{gundlach-blum_2013} as $f_1 = 5.18 \times 10^{-2}, f_2 = 5.26, 
\chi=0.41$. The term $\lambda_{\text{solid}}$ is the thermal conductivity of the solid material of a particle that makes up the regolith. This term generally depends on the composition, and it can be extrapolated from meteorites analogues associated with the asteroid spectral type.
\citet{zhang-etal_2024} proposed that Kamo`oalewa is similar to LL chondrite, therefore results on similar meteorites can be used. \citet{opeil-etal_2010} found a value of conductivity of 1.5 W m$^{-1}$ K$^{-1}$ on a sample of L chondrite, while measurements on LL5 chondrites from the Chelyabinsk meteorite revealed a value of 4.1 W m$^{-1}$ K$^{-1}$ \citep{szurgot-etal_2014}. Since these are the only meteorites most similar to LL chondrite for which data are available, we decided to use three different reference values for the heat conductivity of the solid material, namely $\lambda_{\text{solid}} = 1.5, 3, 4$ W m$^{-1}$ K$^{-1}$.
The values of the Poisson ratio, the Young modulus, and the specific energy of the grain material are determined from measurements on basaltic rocks, and we assumed the values $\mu = 0.25$, $E = 7.8 \times 10^{10}$ Pa, and $\gamma(T) = (6.67 \times 10^{-5} \text{ J m$^{-2}$ K$^{-1}$)} \times T$.
Since Kamo`oalewa has a semi-major axis close to 1 au, we assumed an average surface temperature of 280 K, that corresponds to the subsolar temperature at 1 au.

The method for the grain size estimation consists in keeping the radius $r$ as a free parameter in Eq.~\eqref{eq:gundblum}, and then solving for $r$ the equation $\lambda(r,T,\phi) = K$, where $K$ is the thermal conductivity value estimated by the method of Sec.\ref{ss:TIest}. Note that $K$ is the thermal conductivity of the regolith, thus it depends also on the porosity of the material aggregate.

\section{Results}
\label{s:res}

\subsection{Orbit determination}
\label{ss:ODresults}
All the available observations of Kamo`oalewa at MPC in the new ADES (Astrometry Data Exchange Standard) format were downloaded through the MPC Explorer\footnote{\url{https://data.minorplanetcenter.net/explorer/}} service. The available observations are taken from 13 different stations, and they cover an arc of about 20 years. 
As pointed out above, we replaced the measurements from SDSS of 17 March 2004 with those we reported in Table~\ref{tab:469219_obs}.
Apart from astrometric positions determined in this work, only some observations from the University of Hawaii 88-inch telescope (MPC code T12) and from Pan-STARRS2 reported astrometric uncertainties. Therefore, we used the uncertainties reported by the observers for astrometric positions from T12 and F52, applying a safety term of 0.05 arcsec (see Sec.~\ref{ss:OD}). Since the tracklets from T12 have many observations per night, we also applied the $\sqrt{N/4}$ factor to the reported uncertainties. For the observations taken in this work, we used the measured uncertainties without applying any safety factor, while all the other observations were treated according to the statistical astrometric error model (see Sec.~\ref{ss:OD}). 

\begin{figure}
    \centering
    \includegraphics[width=0.5\textwidth]{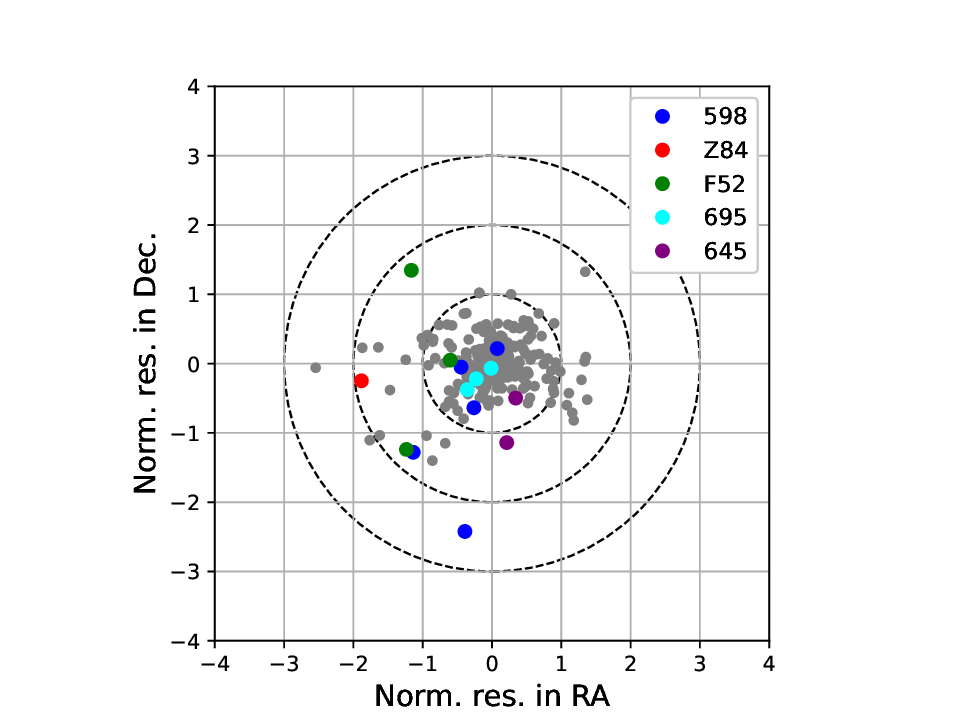}
    \caption{Distribution of the post-fit residuals in RA and Dec., normalized by their post-fit astrometric error. Blue and red dots correspond to observations taken in this work from 568 and Z84, respectively. Green and cyan dots correspond to observations from F52 and 695, respectively, taken in 2024. Violet dots correspond to re-measurements of images from 645. Dashed circles represent values of $\chi$ equal to 3, 2, and 1.}
    \label{fig:residuals}
\end{figure}

All the observations were accepted in the 7-dimensional orbit determination, as their normalized residuals in RA and Dec. all fell within 3$\sigma$ from the assumed astrometric error, see Fig.~\ref{fig:residuals}. The root-mean-square (RMS) error of the normalized residuals resulted to be 0.452. Orbital elements computed with the ESA Aegis software are shown in Table~\ref{tab:orbit}. The new astrometric positions taken in this work permitted to determine the Yarkovsky effect parameter $A_2$ acting on Kamo`oalewa with a value of $(-157.31\pm 11) \times 10^{-15}$ au d$^{-2}$, providing a detection with a SNR of 14.3. 
The corresponding value of semi-major axis drift computed with Eq.~\eqref{eq:A2todadt} amounts to $(-67.35 \pm 4.70) \times 10^{-4}$ au My$^{-1}$.
Without these new observations from 2024, the orbital solution resulted in a Yarkovsky detection with SNR ranging from 2 to 3, depending on the assumptions made in the orbit determination process \citep{liu-etal_2022, hu-etal_2023}.

\begin{figure*}
    \centering
    \includegraphics[width=0.98\textwidth]{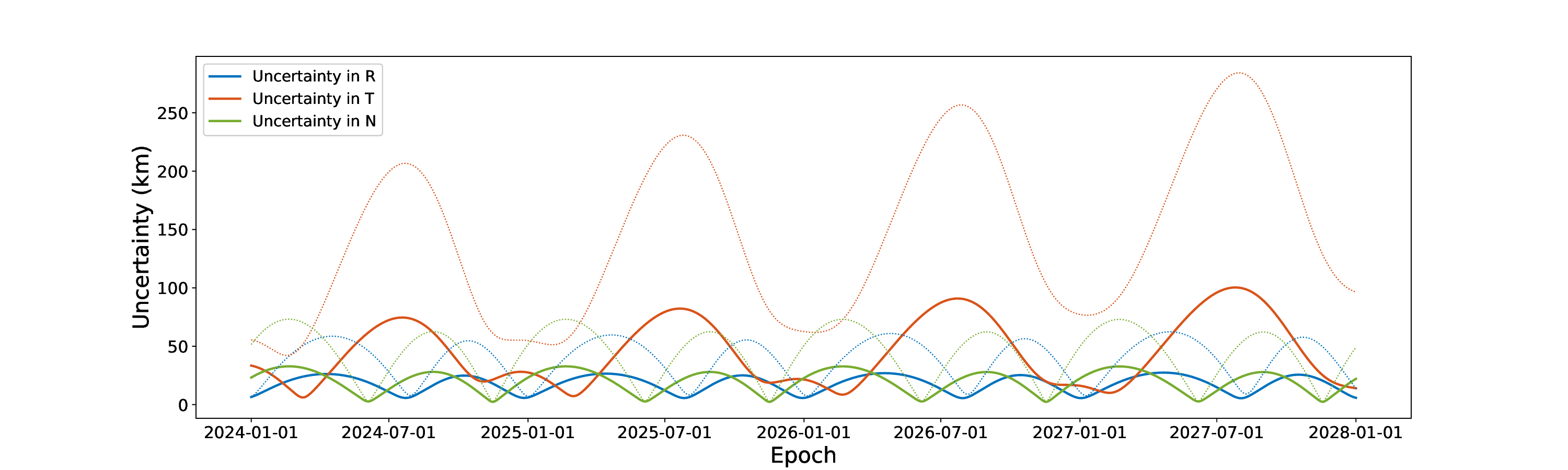}
    \caption{Propagated $1$-sigma uncertainty along radial (blue), transversal (red), and normal to the orbital plane (green) directions, from 1 January 2024 to 1 January 2028. Solid thick curves represent the uncertainties computed with the orbital solution computed here. Dotted curves are the uncertainties computed with the orbital solution which does not use the observations from 2024. }
    \label{fig:OD_uncertainty_propag}
\end{figure*}

The new observations and the new orbit computed here permit to improve the ephemeris of the Kamo`oalewa at the time of the arrival of the Tianwen-2 probe. Propagated 1$\sigma$ uncertainties along radial, transversal, and normal to the orbital plane directions are shown in Fig.~\ref{fig:OD_uncertainty_propag}. We can see that the largest uncertainty is along the transversal direction, and it oscillates between about 20 and about 80 km. The figure also shows the uncertainties in the orbital solution computed without the new observations taken in 2024, proving that the uncertainty dropped significantly with the new orbital solution presented here.
{\renewcommand{\arraystretch}{1.2} 
\begin{table*}[!ht]
    \centering
    \caption{Keplerian orbital elements of (469219) Kamo`oalewa, with the corresponding epoch. Errors refer to the 1-$\sigma$ formal uncertainties.}
    \begin{tabular}{lr}
         \hline
         \hline
      Parameter                    & Value  \\
         \hline
      Epoch                          & 2017-06-17 01:14:57.610 UTC  \\ 
      Semi-major axis, $a$ (au)                &  $  1.0012152878   \pm  1.97	\times 10^{-9}$ \\ 
      Eccentricity, $e$                        &  $  0.1039432071   \pm  1.64	\times 10^{-7}$ \\ 
      Inclination, $i$ (deg)                   &  $  7.7759449225   \pm  1.16	\times 10^{-5}$ \\ 
      Longitude of node, $\Omega$ (deg)        &  $ 66.4282867103   \pm  1.31	\times 10^{-5}$ \\ 
      Argument of perihelion, $\omega$ (deg)   &  $307.0011207190   \pm  1.74	\times 10^{-5}$ \\ 
      Mean anomaly, $\ell$  (deg)            &    $253.7415024247   \pm  2.74	\times 10^{-5}$ \\ 
      $A_2$ ($10^{-15}$ au d$^{-2}$) &  $-157.31 \pm 11.00$                           \\
      $\langle \text{d}a/\text{d}t \rangle$ ($10^{-4}$ au My$^{-1}$) &  $-67.35 \pm 4.70$\\
      Normalized RMS                 & 0.452 \\
         \hline
    \end{tabular}
    \label{tab:orbit}
\end{table*}
}

Knowing that bad astrometry can significantly affect the Yarkovsky effect detection \citep{farnocchia-etal_2013, delvigna-etal_2018, fenucci-etal_2024}, we also performed additional tests to confirm the $A_2$ signal. 
We first used a more aggressive outlier rejection scheme with $\chi^2_{\text{rej}} = 4$. With this additional constraints, we obtained a value of $A_2$ of $(-153.62 \pm 12.65) \times 10^{-15}$ au d$^{-2}$, which is still statistically compatible with the previous nominal solution at 1$\sigma$ level. We also tested the orbit determination by: 1) using weights from statistical data for T12 instead of the ADES weights; 2) using ADES weights only for the observations taken in this work; 3) using the statistical model by \citet{veres-etal_2017} without ADES weights. In case 1) we obtained $A_2 = (-152.60 \pm 11.71) \times 10^{-15}$ au d$^{-2}$, in case 2) $A_2 = (-154.49 \pm 12.52) \times 10^{-15}$ au d$^{-2}$, and in case 3) $A_2 = (-136.28 \pm 15.82) \times 10^{-15}$ au d$^{-2}$. The detection of case 3) is the one with the lowest SNR, but it is still consistent with the others within 1$\sigma$. We therefore consider the $A_2$ determination reported in Tab~\ref{tab:orbit} as a positive and reliable Yarkovsky effect detection. In the next, we assume the semi-major axis drift value of the nominal solution of Tab.~\ref{tab:orbit} for the thermal inertia estimation.

\subsection{Thermal inertia and regolith size estimation}

\subsubsection{Thermal inertia estimation}
\label{res:TI}

\begin{figure*}
    \centering
    \includegraphics[width=0.95\textwidth]{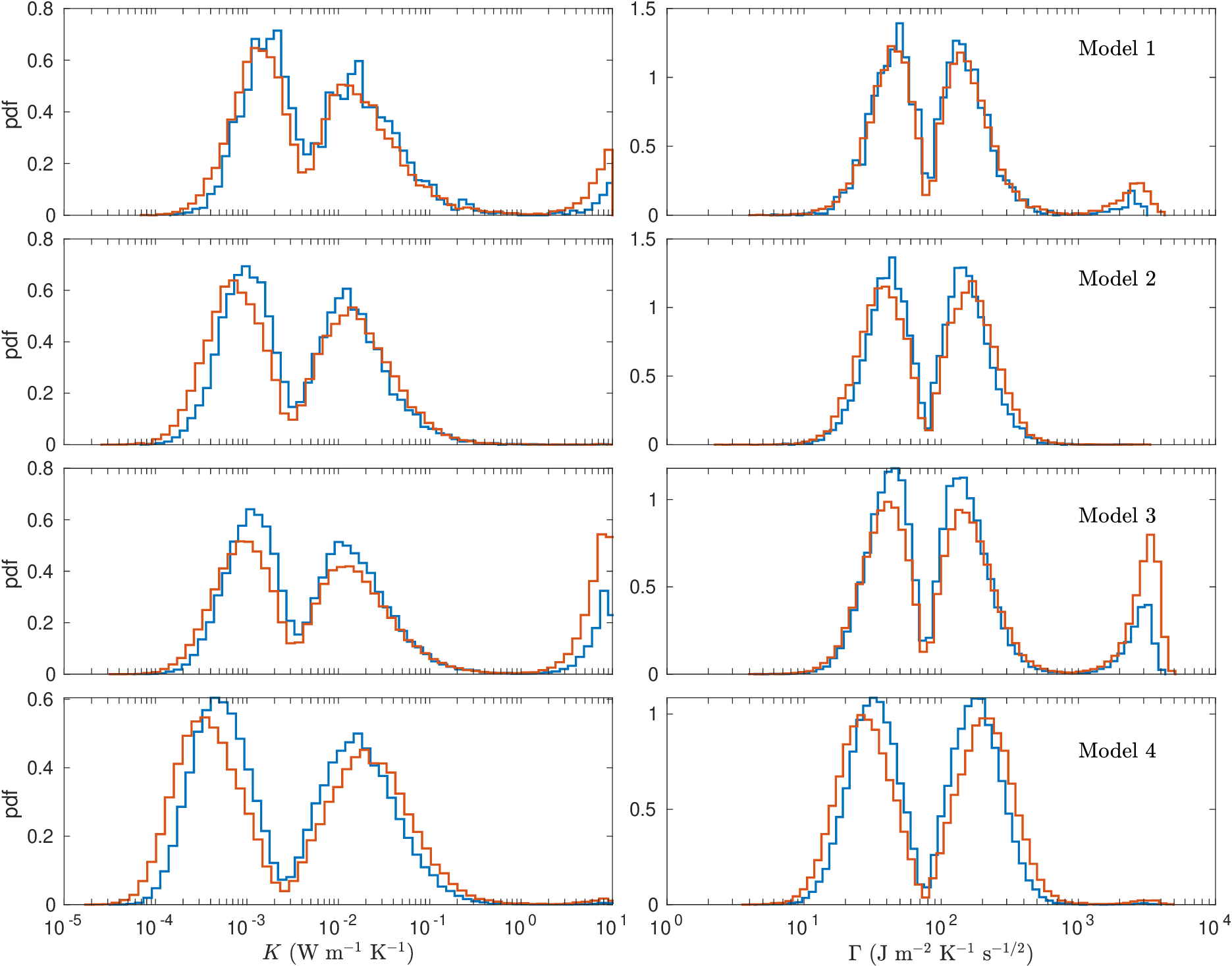}
    \caption{Distribution of thermal conductivity $K$ (left columns) and thermal inertia $\Gamma$ (right column) of Kamo`oalewa, obtained with the ASTERIA method. Different rows correspond to different physical parameter modeling listed in Table~\ref{tab:physPar}. Blue histograms correspond to results obtained with the nominal orbit solution of Tab.~\ref{tab:orbit}. Red histograms are results obtained with the lower SNR Yarkovsky detection of $A_2 = (-136.28 \pm 15.82) \times 10^{-15}$ au d$^{-2}$ which does not take into account the measured astrometric errors for specific observations. }
    \label{fig:TI_distribution}
\end{figure*}

To get an estimate of the thermal inertia, we used the four different physical models described in Table~\ref{tab:physPar}. Distributions of the thermal conductivity $K$ and thermal inertia $\Gamma$ are shown in Fig.~\ref{fig:TI_distribution} for all the different scenarios. All the distributions show either a bimodality or a multimodality with three peaks, which are expected for the ASTERIA method \citep{fenucci-etal_2021, fenucci-etal_2023}.
{\renewcommand{\arraystretch}{1.3} 
\begin{table}[!ht]
    \centering
    \caption{Summary of the thermal inertia results for the different physical models of Kamo`oalewa. The nominal values of the peaks are reported in columns 2, 3, and 4. The best estimate for $\Gamma$ is reported in column 5. All the values are reported with usual thermal inertia units of J m$^{-2}$ K$^{-1}$ s$^{-1/2}$. Column 6 shows the probability that $\Gamma$ is smaller than 1000 J m$^{-2}$ K$^{-1}$ s$^{-1/2}$.}
        \setlength{\tabcolsep}{3pt}
    \begin{tabular}{cccccc}
    \hline 
    \hline
    Model & Peak 1 & Peak 2 & Peak 3 & Best estimate & $P(\Gamma < 1000)$  \\
    \hline
    1 & 46 & 138 & 2456 & $147^{+77}_{-40}$ & 0.97 \\
    2 & 41 & 145 & /    & $152^{+74}_{-42}$ & 0.99 \\
    3 & 43 & 136 & 2947 & $148^{+77}_{-41}$ & 0.91 \\
    4 & 31 & 177 & /    & $181^{+95}_{-59}$ & 0.99 \\
    \hline
    \end{tabular}
    \label{tab:TIresults}
\end{table}
}

The values of the peaks, of the best estimate of thermal inertia, and of the probability that $\Gamma$ is smaller than 1000~J m$^{-2}$ K$^{-1}$ s$^{-1/2}$, are reported in Table~\ref{tab:TIresults}. As can be noted, there is a small variability in the values of the first peak, which are all around 40~J m$^{-2}$ K$^{-1}$ s$^{-1/2}$. The second peak appears at values of about 140~J m$^{-2}$ K$^{-1}$ s$^{-1/2}$, except for Model 4, which corresponds to the high-albedo case with NEO population obliquity. In this case, the second peak is shifted towards slightly larger values of 180~J m$^{-2}$ K$^{-1}$ s$^{-1/2}$. A third peak at values larger than 2000~J m$^{-2}$ K$^{-1}$ s$^{-1/2}$ appears only in Model 1 and Model 3, when the obliquity is given the value of $107 \pm 10$ deg. %

In light of the above results, we can discuss which solution may be the most likely for Kamo`oalewa. 
Low thermal inertia values of $\sim$40~J m$^{-2}$ K$^{-1}$ s$^{-1/2}$ for the first peak correspond to values of conductivity smaller than 0.001 W m$^{-1}$ K$^{-1}$, that are typically too low for NEAs \citep{delbo-etal_2015, novakovic-etal_2024}, and it seems too small to be a likely solution for Kamo`oalewa. 
The high thermal inertia peak at values larger than $2000$ J m$^{-2}$ K$^{-1}$ s$^{-1/2}$ of Model 1 and 3 correspond to solutions with obliquity near 90 deg. At these values of obliquity, the seasonal component of the Yarkovsky effect either has the same magnitude or dominates over the diurnal effect, and a high thermal inertia is necessarily needed to slowly re-radiate the heat in order to achieve the high Yarkovsky drift that we determined from astrometry. However, these configurations have a low probability to occur of about 0.05 and 0.13 for Model 1 and Model 3, respectively. 
Values of thermal inertia $>$2000 J m$^{-2}$ K$^{-1}$ s$^{-1/2}$ are typical of compact bare rocks and closer to measurements obtained on meteorites \citep[see][for a review]{ostrowsky-bryson_2019}, which typically have a much lower porosity than asteroids.
In addition, the values of the density associated with the high thermal inertia solutions are all smaller than 1700 kg m$^{-3}$. 
These values of density do not appear compatible with the thermal inertia estimate and the known physical properties of Kamo`oalewa. In fact, density of 1700 kg m$^{-3}$ and thermal inertia larger $2000$ J m$^{-2}$ K$^{-1}$ s$^{-1/2}$ could be achieved by a monolithic block with a porosity below 10\%. However, this would imply a primitive C-type-like composition, which is not compatible with the known spectrum of Kamo`oalewa. 
On the other hand, if we assume that Kamo`oalewa has a S-type composition, a density of 1700 kg m$^{-3}$ would imply a high porosity, that in turn reduces the thermal inertia value \citep[see e.g.][]{okada_2016}. 
Further, if we look at the values of the conductivity $K$, the third peak corresponds to values of 7.5 W m$^{-1}$ K$^{-1}$, and this is generally not compatible with density smaller than 2000 kg m$^{-3}$ \citep{soini-etal_2020}.
Taking into account these considerations, the third peak at high inertia values seems very unlikely. 

It is therefore most likely that the actual thermal inertia value of Kamo`oalewa falls within the second peak of the distribution, which we consider as the nominal estimate of each model. Taking into account the spread of the distribution, the estimates obtained with Model 1, 2, and 3 are all very similar to each other, and within statistical fluctuations they give a value of $\Gamma = 150^{+90}_{-45}$ J m$^{-2}$ K$^{-1}$ s$^{-1/2}$. The corresponding most likely value of thermal conductivity $K$ is in the range 0.012 - 0.015 W m$^{-1}$ K$^{-1}$. 
Model 4 on the other hand gives a slightly higher estimate at $\Gamma = 181^{+95}_{-60}$ J m$^{-2}$ K$^{-1}$ s$^{-1/2}$. 
This difference arises because Model 4, which assumes high albedo and a population-based obliquity distribution, favors smaller diameters and obliquities around 140°. In contrast, Models 1, 2, and 3 either favor larger sizes due to lower albedo and/or an obliquity closer to 90 deg ($107 \pm 10$ deg), both of which reduce the efficiency of the diurnal Yarkovsky effect. As a result, Model 4 allows the measured drift to be reproduced even with a slightly higher thermal inertia.

\begin{figure*}
    \centering
    \includegraphics[width=0.9\textwidth]{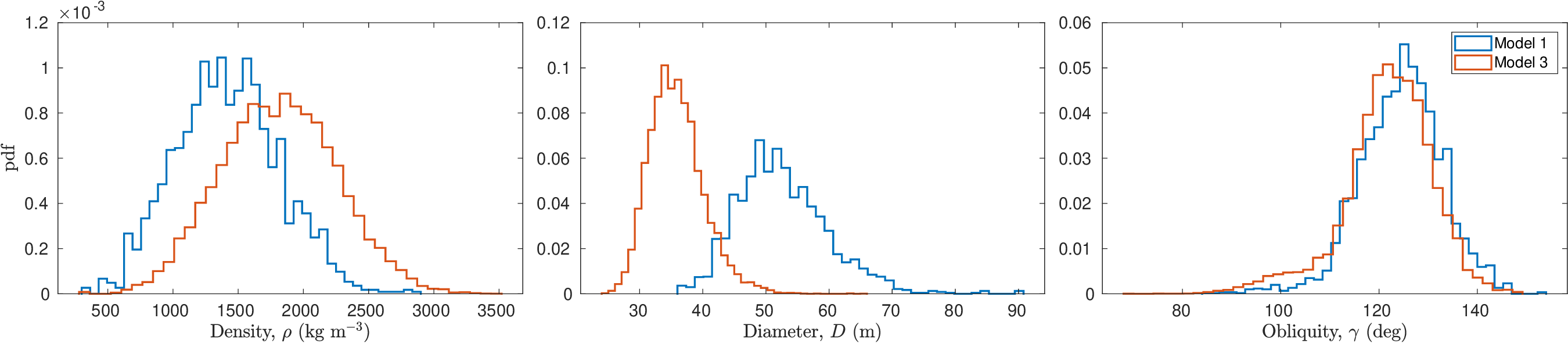}
    \caption{Output distribution of density (left panel), diameter (central panel), and obliquity (right panel) obtained for Model 1 (blue) and Model 3 (red).}
    \label{fig:Kamooalewa_other_params}
\end{figure*}

Despite the estimate of thermal inertia is very similar in all the cases, taking into account the value of the obliquity in Model 1 and Model 3 puts also additional constraints on other parameters of the model, because Eq.~\eqref{eq:asteria_eq} can not be solved for any combination of input parameters. Figure~\ref{fig:Kamooalewa_other_params} shows the output distributions of density, diameter, and obliquity. The density is skewed towards smaller values than what assumed in input, with a median value of about 1500 kg m$^{-3}$ for Model 1 and of 1800 kg m$^{-3}$ for Model 3. For the obliquity, values larger than 110 deg are favored by the model, with a most likely value at about 120 deg. The diameter distribution resembles the assumed input, although it is slightly shifted towards slightly smaller values in both cases.

\subsubsection{Grain size estimation}
\begin{figure*}
    \centering
    \includegraphics[width=0.48\textwidth]{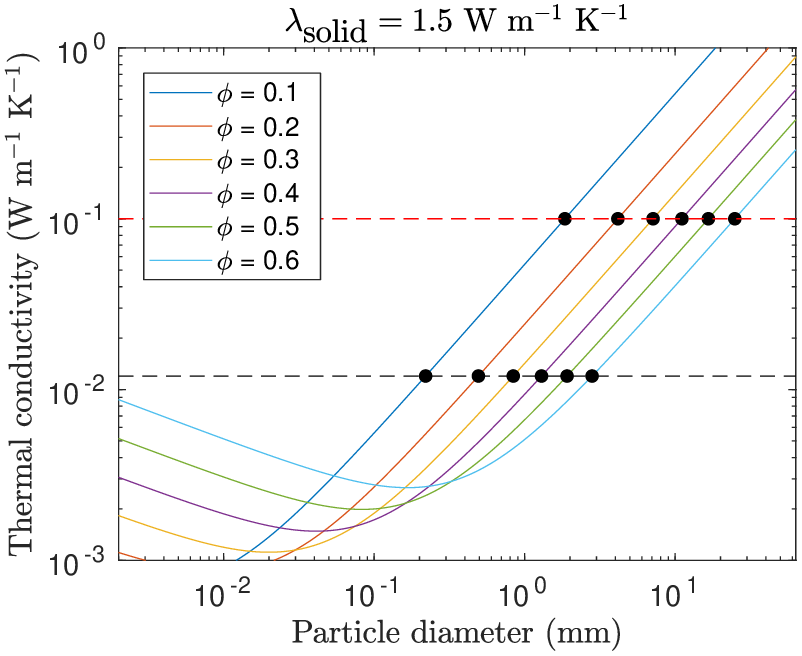}
    \includegraphics[width=0.48\textwidth]{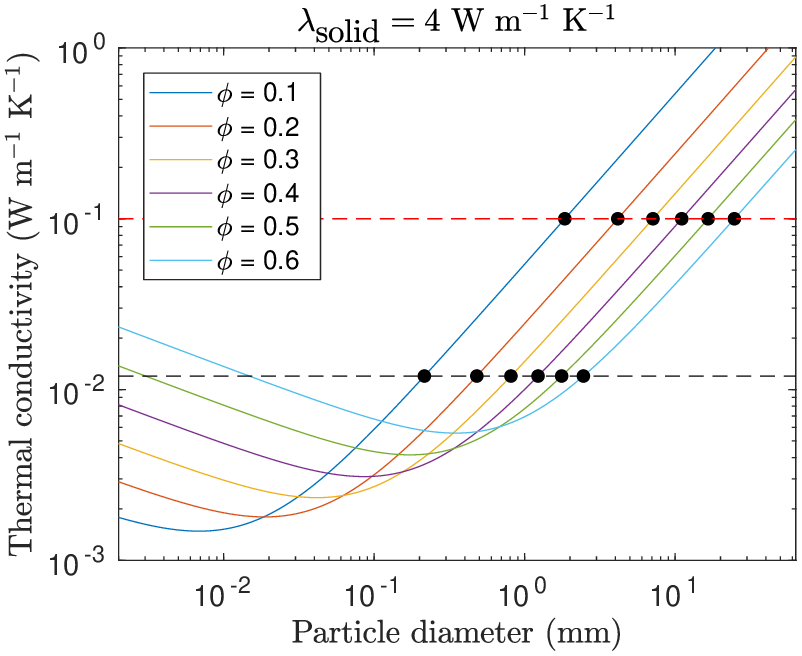}
    \caption{Regolith particles diameter for different filling factors $\phi$. The horizontal dashed black line corresponds to the nominal estimate of $K$ of $0.012$ W m$^{-1}$ K$^{-1}$, the dashed red line to the upper limit of $0.1$ W m$^{-1}$ K$^{-1}$, and the colored curves depict the dependency of the thermal conductivity as a function of the regolith particle diameter. The conductivity of solid material was assumed to be $\lambda_{\textrm{solid}} = 1.5$ W m$^{-1}$ K$^{-1}$ for the plot on the left, and $\lambda_{\textrm{solid}} = 4$ W m$^{-1}$ K$^{-1}$ for the plot on the right. The intersections between the horizontal dashed lines and the curves, indicated by black circles, correspond to the estimated grain size. }
    \label{fig:grain_size}
\end{figure*}

The estimated thermal conductivity $K$ was used to estimate the size of regolith eventually present on the surface by using the method of Sec.~\ref{ss:grainsize}. Values of $K$ in the range 0.012 - 0.015 W m$^{-1}$ K$^{-1}$ corresponding to the second peak of the distribution were used for the computations. As a reference, we also computed the grain size for the upper limit $K$ value of 0.1 W m$^{-1}$ K$^{-1}$, which is at the low-probability tail of the distributions of Fig.~\ref{fig:TI_distribution}. We present the results obtained assuming $\lambda_{\textrm{solid}} = 1.5$ W m$^{-1}$ K$^{-1}$ and $\lambda_{\textrm{solid}} = 4$ W m$^{-1}$ K$^{-1}$. Filling factor values of $\phi = 0.1, \dots, 0.6$ with step $\Delta \phi = 0.1$ were used, since it is not possible to get an estimate for the porosity of the surface material. 
Figure~\ref{fig:grain_size} shows the thermal conductivity curves modeled with Eq.~\eqref{eq:gundblum}, and the horizontal dashed lines correspond to the nominal value $K = 0.012$ W m$^{-1}$ K$^{-1}$ (black dashed) and the upper limit value $K = 0.1$ W m$^{-1}$ K$^{-1}$ (red dashed). 
The particle diameter value is obtained at the intersection between the estimated $K$ value and that modeled from Eq.~\eqref{eq:gundblum}. 
Note that here we give results about particle diameter, whereas the model presented in Sec.~\ref{ss:grainsize} takes into account the particle radius. Regolith of size between $\sim$0.1 and 3 mm are compatible with the estimated thermal conductivity, and the main source of uncertainty is given by the unknown filling factor. 
For the upper limit thermal conductivity, the particle size reaches diameters as big as $\sim$13 mm.

For the reasons explained in Sec.~\ref{ss:grainsize}, we also used different values of thermal conductivity of solid material $\lambda_{\textrm{solid}}$. 
As can be seen, changing the conductivity of the solid material to $\lambda_{\textrm{solid}} = 4$ W m$^{-1}$ K$^{-1}$ did not significantly change in the results, and the same applies for $\lambda_{\textrm{solid}} = 3$ W m$^{-1}$ K$^{-1}$. This is because, for small asteroids, the radiative heat transport term of Eq.~\eqref{eq:gundblum} is dominating in the estimation. 

Note that this approach based on \citet{gundlach-blum_2013} necessarily assumes that a fine regolith layer is present on the surface of the asteroid. As previous missions on Bennu and Ryugu demonstrated, extrapolating regolith size based on the measurement of thermal inertia is not always straightforward \citep{lauretta-etal_2019, watanabe-etal_2019, rozitis-etal_2020}, thus our particle size estimate should be taken with caution. As demonstrated by \citet{ryan-etal_2022}, regolith particle size forecasts based on the thermal inertia are mainly independent of assumptions on regolith porosity, except for when the non-isothermality effect is considerable, as is the case when the regolith is notably coarse and/or composed of high microporosity materials.

\section{Discussion}
\label{s:discussion}

\subsection{Sensitivity to input parameters}
In Sec.~\ref{ss:ODresults}, we showed that the Yarkovsky effect determination lowers to $A_2 = (-136.28 \pm 15.82) \times 10^{-15}$ au d$^{-2}$ if astrometric errors are all set according to \citet{veres-etal_2017}, which corresponds to a semi-major axis drift of $\textrm{d}a/\textrm{d}t = (-77.76 \pm 9.4) \times 10^{-4}$ au My$^{-1}$. Thermal inertia estimates using the same models of Tab.~\ref{tab:physPar} and this lower SNR Yarkovsky effect determination are presented in Fig.~\ref{fig:TI_distribution} with red histograms. Overall, the distributions obtained here and in Sec.~\ref{res:TI} are similar, with values of the second peak shifted towards slightly larger values of thermal inertia for the low SNR Yarkovsky determination. We also note that the third peak at high thermal inertia has a higher probability. Both these effects are explained by the fact that the lower SNR determination allows smaller semi-major axis drifts, which can be achieved by a higher thermal inertia. Still, the solutions at around $~$150 or $\sim$180 J m$^{-2}$ K$^{-1}$ s$^{-1/2}$ are the ones preferred also in this case.

The brightness of Kamo`oalewa increases to about $V=22$ mag periodically between February and April every year, therefore there are chances to perform further astrometry before the arrival of the Tianwen-2 spacecraft. The predicted sky-plane uncertainty in this time window in 2025 and 2026 is lower than the typical level of astrometric accuracy; therefore the SNR in the $A_2$ parameter is unlikely to improve again with only ground-based optical observations. Despite this, we still tested how a better SNR in the Yarkovsky effect determination changes the results. To this purpose, we used the nominal value of $\text{d}a/\text{d}t$ reported in Table~\ref{tab:orbit}, and artificially assigned an SNR of 100. As a test, we produced results only for Model 2. In this way, we obtained a thermal inertia of $\Gamma = 153^{+73}_{-42}$ J m$^{-2}$ K$^{-1}$ s$^{-1/2}$, which is statistically compatible with the results obtained in the previous sections. This indicates that, at current level of knowledge of the physical and orbital properties of Kamo`oalewa, the uncertainties in the thermal inertia are mainly governed by the least constrained physical parameters. Therefore, to improve thermal inertia estimation from ground-based observations only, efforts in determining other quantities rather than the Yarkovsky effect should be pursued.  

Another test that we performed is about the Yarkovsky effect modeling. Results for asteroid Didymos by \citet{novakovic-fenucci_2024} showed that thermal inertia estimations obtained with the ASTERIA model may be sensitive to variable thermal inertia along the orbit. The current model implemented in ASTERIA assumes that $\Gamma$ varies with the distance $r$ from the Sun as $\Gamma_0 r^{\beta}$, where $\Gamma_0$ is the thermal inertia at 1 au and $\beta$ is a free parameter. The expected exponent for radiative heat transfer is $\beta = -0.75$, however \citet{rozitis-etal_2018} showed that it can be as large as $\beta = -2$. Although the eccentricity of the orbit of Kamo`oalewa is about 0.1, results obtained in \citet{fenucci-etal_2021} showed that variable thermal inertia may affect the results also for moderately eccentric orbits. To test this, we performed additional simulations with values of $\beta$ of $-0.75, -1$ and $-2$ still with Model 2. We did not find any statistically significant difference in the thermal inertia estimation, which resulted to be $\Gamma = 151^{+79}_{-46}$ J m$^{-2}$ K$^{-1}$ s$^{-1/2}$ with only small variations on the nominal value.
\begin{figure*}
    \centering
    \includegraphics[width=0.95\textwidth]{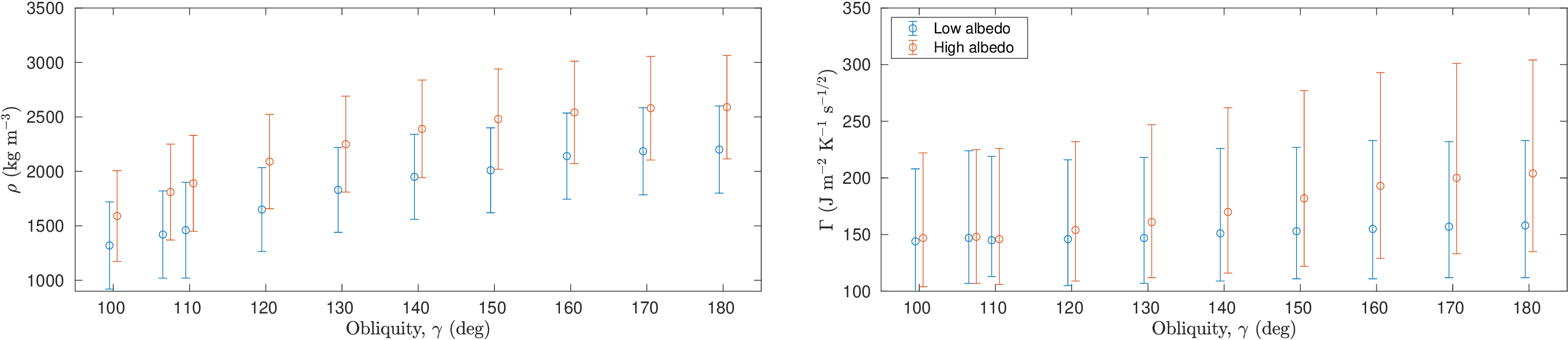}
    \caption{Output values for the density (left panel) and for the thermal inertia (right panel), obtained for input Gaussian distributions of obliquity with values $\gamma = (100 + 10k) \pm 10$ deg, with $k = 1,\dots,8$, for the low-albedo Model 1 (blue markers) and the high-albedo case Model 3 (red markers).}
    \label{fig:Kamooalewa_gamma_dependence}
\end{figure*}

On the other hand, results obtained in Sec.~\ref{s:res} showed that both thermal inertia and the output density are somewhat sensitive to the value of the obliquity. To further illustrate this, we performed additional simulations with different Gaussian distributions of the obliquity as $\gamma = (100 + 10k) \pm 10$ deg, for $k=0,\dots,8$. We performed the computations for both cases of low and high albedo described in Sec.~\ref{s:methods}.
Figure~\ref{fig:Kamooalewa_gamma_dependence} shows the distribution of both density and the second peak of thermal inertia, as a function of the input obliquity. The thermal inertia shows small changes in the estimated nominal value for the case of low albedo, and all the distributions do not show statistically significant differences overall. For the high albedo case, the nominal value shifts towards slightly larger values, and the uncertainty increases reaching values up to 300 J m$^{-2}$ K$^{-1}$ s$^{-1/2}$ for obliquity of 180 deg. 
On the contrary, there is a clear trend in the output density distribution. The median value varies from about 1500 kg m$^{-3}$ for obliquity close to 100 deg, up to 2350 kg m$^{-3}$ for obliquity close to 180 deg for the low albedo case. In the high albedo case, the median value at 100 deg in obliquity is about 1800 kg m$^{-3}$, and it reaches 2500 kg m$^{-3}$ at obliquity of 180 deg. 
This trend is determined by the fact that, as the obliquity increases, the diurnal Yarkovsky effect becomes more efficient. On the other hand, increasing the density reduces the Yarkovsky drift, because the heat re-radiation should move a larger mass. Since the measured drift in Eq.~\eqref{eq:asteria_eq} remains constant, the increase in the diurnal Yarkovsky effect caused by increasing the obliquity is compensated by an increase of the density, so that the overall drift is kept to the measured value.
Although there is already a first determination of the obliquity from \citet{zhang-etal_2025}, confirming the estimated value, along with a determination of the albedo, would be of extreme importance to better characterize and improve the physical properties estimation of Kamo`oalewa prior to the arrival of the Tianwen-2 spacecraft. To this purpose, additional lightcurves could be obtained at the next opposition, and other lightcurve inversions could be attempted in order to determine the obliquity.  

\subsection{Role of surface roughness}

A limitation of our model is that it does not include surface roughness, which is an important factor influencing the temperature distribution of airless bodies, including asteroids \citep[e.g.][]{muller-etal_2014}. This is particularly relevant for modeling the Yarkovsky effect \citep[see, e.g.][]{2021Icar..36914594F}. Using the Advanced Thermophysical Model (ATPM), \citet{2011MNRAS.415.2042R,2012MNRAS.423..367R} showed that thermal-infrared beaming due to surface roughness can enhance the Yarkovsky-induced orbital drift by tens of percent and, in some cases, by up to a factor of two.  

However, as noted by \citet{fenucci-etal_2021}, our Yarkovsky model assumes nonlinear boundary conditions for heat conduction. While surface roughness tends to amplify the Yarkovsky effect, the nonlinearity of heat conduction generally results in a slight reduction in the predicted Yarkovsky-induced drift of the semi-major axis. As a result, these two effects partially counterbalance each other.  

Next, despite the high accuracy of our Yarkovsky drift detection, the knowledge of Kamo`oalewa's key parameters remains limited. Specifically, uncertainties in its bulk density, size, and spin axis orientation contribute to variability in our model, and are likely more significant than the uncertainty introduced by neglecting surface roughness.

Additionally, separating thermal inertia and surface roughness is impossible within our model because both parameters similarly influence temperature distributions and infrared emissions. High surface roughness can mimic high thermal inertia by trapping heat in concavities and releasing it over an extended period. As a result, when the thermal inertia of an asteroid's surface is estimated assuming a smooth surface (no roughness), the actual thermal inertia is likely lower than the estimated value. Since our results for Kamo`oalewa already suggest relatively low thermal inertia, incorporating surface roughness into our analysis would not change our overall conclusions — it would only slightly further reduce the estimated thermal inertia.  

\subsection{Geological implications}
Thermal inertia for sub-km S-type NEAs is available only for a handful of objects \citep{novakovic-etal_2024}. Among them, the only one visited by a spacecraft is Itokawa, which has a thermal inertia of $700 \pm 200$ J m$^{-2}$ K$^{-1}$ s$^{-1/2}$ \citep{muller-etal_2014}, which is well above the values obtained here for Kamo`oalewa. 
In contrast, the range of values that we estimated is closer to those obtained for Bennu and Ryugu from in-situ mission explorations, which are $320 \pm 30$ and $220 \pm 45$ J m$^{-2}$ K$^{-1}$ s$^{-1/2}$ \citep{rozitis-etal_2020, shimaki-etal_2020}, respectively. However, Kamo`oalewa does not show a spectrum similar to that of C-type asteroids, and it is not possible to draw a conclusion about the surface composition by a comparison with them. In contrast, the low value of thermal inertia estimated here are more similar to those of larger asteroids, and this may be an indication of the presence of coarse regolith on the surface of Kamo`oalewa, that is generally able to create a thermal insulating layer on the surface.
By studying the stability of eventual small grains laying on the surface of Kamo`oalewa, \citet{zhang-etal_2024} found that the critical regolith size that can withstand the fast rotation rate is of the order of $\sim$40 mm near the pole regions, and of $\sim$5 mm around the equatorial region. Similar results were also found in a previous work by \citet{li-scheeres_2021}.
The range $\sim$$0.1 -3$ mm of regolith size obtained in Sec.~\ref{res:TI} from the thermal conductivity estimation is completely compatible with the maximum regolith size estimated by \citet{zhang-etal_2024}, and grains on this size can remain on the whole surface of Kamo`oalewa without getting expelled by the fast rotation. Therefore, it may be possible that Kamo`oalewa is a fast-rotator completely covered by coarse regolith grains.

For the case of Ryugu, laboratory analysis of the sample returned on Earth showed that their thermal inertia is a few times larger than that measured in-situ \citep{ishizaki-etal_2023}. Recently, \citet{hamm-etal_2024} proposed that the low thermal inertia measured in-situ on Ryugu could be due to fracturing of boulders with high bulk thermal inertia, which would cause the average thermal inertia to drop down. Note that the fracturing process due to thermal stress was proposed by \citet{delbo-etal_2014}, and confirmed by in-situ observations on Bennu \citep{molaro-etal_2020}. Fracturing of the surface could also then be an explanation for the low thermal inertia estimated for Kamo`oalewa. 

Another scenario that could account for the low thermal inertia is that of high micro-porosity of material. Rocks with high micro-porosity and low thermal inertia have been found on both Ryugu and Bennu \citep{grott-etal_2019, ryan-etal_2024}, and they were proven not to hold fine dust on their surface. Thus, even if Kamo`oalewa is not a carbonaceous asteroid as Ryugu and Bennu, it may still have a similar internal structure and thermal properties. We could thus hypothesize that even S-type monolithic blocks might form with significant micro-porosity. The in-situ exploration of Kamo`oalewa by the Tianwen-2 mission and the laboratory analysis of the sample eventually returned to Earth will help in clarifying whether this hypothesis holds also for S-type asteroids or not. 

The last hypothesis that could account for the low thermal inertia is that Kamo`oalewa has a significant degree of macro-porosity, thus implying it has a rubble-pile structure. The porosity can be estimated as $\mathcal{P} = (1 - \rho/\rho_{\textrm{m}})$ \citep{carry_2012}, where $\rho_m$ is the bulk density of the meteorite associated with the asteroid. Since \citet{zhang-etal_2024} identified the LL-chondrite as associated meteorite to Kamo`oalewa, we used a value of $\rho_{\textrm{m}} = 3200$ kg m$^{-3}$ \citep{ostrowsky-bryson_2019}. Similarly, assuming that the estimated obliquity of around 107 deg is correct, the output distribution of Kamo`oalewa's density shown in Fig.~\ref{fig:Kamooalewa_other_params} applies. This leads to a porosity of $\mathcal{P}=0.53^{+0.13}_{-0.16}$ for the low-albedo Model 1 and of $\mathcal{P}=0.42^{+0.17}_{-0.17}$ for the high-albedo Model 3, which may be high enough to explain the low thermal inertia. 
However, a rubble-pile structure seems less likely for Kamo`oalewa. The rotation period of 28 minutes is much shorter than the disruption limit of 2.2 hours for cohesion-less rubble-piles \citep{walsh_2018}, therefore some cohesion would be needed in order to prevent the asteroid from disrupting \citep[see e.g.][]{rozitis-etal_2014}. 
According to results presented in \citet{hu-etal_2021}, the minimum cohesion needed for a 50 meters asteroid to withstand a rotation period of 27 minutes is about only 10 Pa, therefore even low values of cohesion may be enough to keep a rubble-pile structure for Kamo`oalewa.
Still, even in the presence of a certain level of cohesion, the long term action of the Yarkovsky–O'Keefe–Radzievskii–Paddack (YORP) effect could lead the asteroid to enter a disaggregation phase, which breaks up the rubble-pile asteroid into its fundamental components in a relatively short time \citep{scheeres_2018}.
The YORP timescale for Kamo`oalewa has been estimated to be between $10^4$ year and $10^5$ year \citep{fenucci-etal_2021}. Note that this timescale is significantly smaller than the typical migration time from the main belt, and of the age of the Giordano Bruno crater \citep[about 5-10 My, see][]{basilevsky-head_2012P} in case Kamo`oalewa originated from the Moon \citep{jiao-etal_2024}. 
In addition to this, \citet{zhang-etal_2024} proposed that Kamo`oalewa underwent a space-weathering exposure of about 100 My in order to show the current high red spectral slope. Thus, if Kamo`oalewa was a rubble-pile, the YORP effect would spin-up the asteroid several times, possibly re-shaping and re-surfacing it \citep{walsh-etal_2012}. Unweathered fresh material would then be brought to the surface, and space-weathering would not have enough time to act to increase the spectral slope to the degree which is observed today.

\section{Conclusions}
\label{s:conclusions}
In this paper, we presented new astrometric measurements of the small NEA (469219) Kamo`oalewa, taken in March 2024 from the Loiano Astronomical Station and from the Calar Alto Observatory. We also re-analyzed and accurately re-measured two precovery images taken by SDSS in March 2004. The major issue in SDSS images is determining the correct observation time, and the discussion presented here is generally valid for SDSS detections. 
We performed the orbit determination of (469219) Kamo`oalewa by using the new observations presented here, plus some additional follow-up observations taken between March and February 2024, and we determined the non-gravitational $A_2$ parameter with an SNR of 14. The new orbit determination also reduced the uncertainty in the position of Kamo`oalewa at the expected time of arrival of the Tianwen-2 mission, setting a largest $1\sigma$ uncertainty of about 80 km. 

Taking advantage of the new determination of the Yarkovsky effect, we used the ASTERIA model to estimate the thermal inertia of the asteroid. 
We used four different ad-hoc models of the physical parameters of Kamo`oalewa, which take into account the different uncertainties in the known properties. 
The ASTERIA model predicted a thermal inertia of $\Gamma = 150^{+90}_{-45}$ J m$^{-2}$ K$^{-1}$ s$^{-1/2}$ or $\Gamma = 181^{+95}_{-60}$ J m$^{-2}$ K$^{-1}$ s$^{-1/2}$, depending on the model used.

These values could be induced by the presence of a regolith layer with average grain size in the range $\sim$0.1 and 3 mm. Results are stable with respect to variations of the physical parameters in a reasonable range of values. On the other hand, the output distribution of the density is somewhat sensitive to the obliquity value, with values close to 100 deg implying a density of about 1500 kg m$^{-3}$. Additional lightcurve data and lightcurve inversion to confirm the estimated obliquity may help to improve the physical characteristics of Kamo`oalewa prior to the arrival of the Tianwen-2 mission. 

Finally, we provided some clues on the structure of Kamo`oalewa based on the results obtained for thermal inertia and density. The low thermal inertia may be caused either by the presence of a regolith layer on the surface, which may withstand the fast rotation, or by internal fracturing. On the other hand, a rubble-pile structure seems unlikely when all the evidence obtained from remote observations are gathered together.

\begin{acknowledgements}
BN acknowledges support by the Science Fund of the Republic of Serbia, GRANT No 7453, Demystifying enigmatic visitors of the near-Earth region (ENIGMA). AC acknowledges the use of the 1.52m Cassini Telescope, run by INAF-OAS Astrophysics and Space Science Observatory Bologna at Loiano. We also thank the anonymous referee whose comments helped us to improve the quality of the manuscript.
\end{acknowledgements}

\bibliographystyle{aa}
\bibliography{holybib.bib}{}

\end{document}